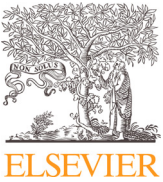
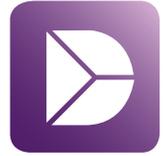

Data Article

# An exploratory assessment of a multidimensional healthcare and economic data on COVID-19 in Nigeria

Ezekiel Ogundepo[1,*], Sakinat Folorunso[2,*], Olubayo Adekanmbi[1], Olalekan Akinsande[1], Oluwatobi Banjo[2], Emeka Ogbuju[3], Francisca Oladipo[3], Olawale Abimbola[1], Ehizokhale Oseghale[1], Oluwatobi Babajide[1]

[1] Data Science Nigeria (DSN)
[2] The Department of Mathematical Sciences, Olabisi Onabanjo University, Ago Iwoye, Ogun State, Nigeria
[3] The Department of Computer Science, Federal University Lokoja, Nigeria

**a b s t r a c t**

The coronavirus disease of 2019 (COVID-19) is a pandemic that is ravaging Nigeria and the world at large. This data article provides a dataset of daily updates of COVID-19 as reported online by the Nigeria Centre for Disease Control (NCDC) from February 27, 2020 to September 29, 2020. The data were obtained through web scraping from different sources and it includes some economic variables such as the Nigeria budget for each state in 2020, population estimate, healthcare facilities, and the COVID-19 laboratories in Nigeria. The dataset has been processed using the standard of the FAIR data principle which encourages its *findability, accessibility, interoperability, and reusability* and will be relevant to researchers in different fields such as Data Science, Epidemiology, Earth Modelling, and Health Informatics.

* Corresponding authors.
*E-mail addresses:* ezekiel@datasciencenigeria.ai, ogundepoezekiel@gmail.com (E. Ogundepo), sakinat.folorunso@oouagoiwoye.edu.ng (S. Folorunso).
*Social media:* (E. Ogundepo).

**Specifications Table**

| | |
|---|---|
| Subject | Infectious diseases, Time series, Econometrics, Epidemiology, Decision science, Data science |
| Specific subject area | A data on the multidimensional study of the effect of COVID-19 on Government budgets, healthcare facilities, and COVID-19 testing laboratories in Nigeria. |
| Type of data | Tables, Charts, and Maps |
| How data were acquired | Data were collected from different sources such as NCDC, Geo-Referenced Infrastructure and Demographic Data for Development (GRID3), Geographic Population and Demographic Data (GeoPoDe), and National Bureau of Statistics (NBS). Some data were acquired directly from the host website using the Google Sheets function IMPORTHTML(). The daily updates of COVID-19 in Nigeria were also scrapped from NCDC timeline on Twitter (twitter.com/NCDCgov) using the rtweet package in R programming [15]. |
| Data format | The raw datasets are in Microsoft Excel format and is available on the Mendeley data repository. |
| Parameters for data collection | Parameters for data collection comprises: <br> 1. Geographical coordinates of all the 36 states and the Federal Capital Territory (FCT). <br> 2. Population estimates of 0 to 100 years per state and by gender. <br> 3. A cumulative report of COVID-19 cases in Nigeria per state. <br> 4. Initial budget presented and the revised budget due to COVID-19 per state in Nigeria. <br> 5. The state initial budget and the revised budget due to COVID-19. <br> 6. Total healthcare facilities in Nigeria per state. <br> 7. COVID-19 laboratories per state. <br> 8. Daily updates of COVID-19 in Nigeria. <br> 9. COVID-19 by the six (6) geopolitical zones in Nigeria. <br> 10. NCDC daily tweets and retweets. |
| Description of data collection | The description of data collection is as follows: <br> ■ The COVID-19 related dataset which included cases by State and laboratories were collected from the NCDC website [3] while daily COVID-19 cases in Nigeria were collected from NCDC Twitter updates [4]. <br> ■ The initial and revised states' budget due to COVID-19 were collected from various news channels in Nigeria [5]. <br> ■ The population estimate per state and by gender were collected from the GeoPoDe [8]. <br> ■ The healthcare facilities per state were collected from GRID3 [6]. <br><br> We used various tools and techniques to build and organise these datasets in a spreadsheet as advised by [16] and is available on the Mendeley data repository. |
| Data source location | Online at http://dx.doi.org/10.17632/8h5rtbbx7m.1 |
| Data accessibility | Repository name: Mendeley <br> Data Repository: http://dx.doi.org/10.17632/8h5rtbbx7m.1 <br> Project URL: https://bit.ly/COVID-19data_project_repo |

**Value of the Data**

- This data explains the effect of COVID-19 on the economy of Nigerian States in terms of budget reallocation and adjustment, and it will be useful for multidimensional studies while carrying out any post-COVID research on Nigeria by government agencies, international organizations and individual researcher.

- The data can help the government of Nigeria prepare the economy better for future pandemics (if any), and it serves as a basis for the appraisal of state government's performance at the end of the fiscal year.
- Data scientists can get hidden information and detect novel patterns from the data which can then be used to train different machine learning models to predict the future of COVID-19 in Nigeria.
- Public health services and disease control institutions can use the data for effective planning using its information on the facilities related to COVID-19 in the Nigerian States.
- It is also relevant for qualitative research works on expert opinion mining on the outbreak to determine both people and government's sentiments on different control measures.

## 1. Data Description

The World Health Organisation (WHO) announced Coronavirus (COVID-19) as a Public Health Emergency of International Concern on January 30, 2020 and a Pandemic on March 11, 2020 [1]. The COVID-19 pandemic pushed the global economy into a Great Lockdown, schools were shut down, travels banned, social distancing enforced, and many jobs were lost. The rate of the spread of the virus keeps increasing, yet no defined countermeasure, remedy or well-tested medications are handy for its eradication. Nigeria recorded her first case of COVID-19 on Thursday, February 27, 2020, and ever since there has been an exponential growth and spread of the virus all over the country. Due to COVID-19 pandemic, there have been some negative changes in the economy of Nigeria such as the total/partial lockdown, banning of international flights, decrease in the national revenue by over N320 billion, a downward revision of each state's budget, building of isolation centers, loss of jobs, and shutting down of schools [2]. This article presents datasets that are related to COVID-19 as reported by the NCDC, healthcare facilities, and laboratories for testing it, and population estimate by gender and budget for all the 36 states of Nigeria and the Federal Capital Territory (FCT). The datasets are in the Microsoft Excel Workbook with five (5) sheets as presented in Fig. 1.

**Sheet 1** This sheet comprises the following features for each state in Nigeria: geocoordinates, population by gender, COVID-19 cases, healthcare facilities, COVID-19 Government laboratories, budget, and total available revenue in 2019.

**Sheet 2** This sheet comprises daily cases of COVID-19 in Nigeria as reported by NCDC on https://twitter.com/NCDCgov.

**Sheet 3** This sheet shows the aggregate of COVID-19 cases in the six (6) geopolitical zones in Nigeria which includes North-Central (Benue, Kogi, Kwara, Nasarawa, Niger, Plateau, and the FCT), North-East (Adamawa, Bauchi, Borno, Gombe, Taraba, Yobe), North-West (Jigawa, Kaduna, Kano, Katsina, Kebbi, Sokoto, Zamfara), South-East (Abia, Anambra, Ebonyi, Enugu,

Fig. 1. A glimpse of the data.

**Table 1**
Description of variables.

| Variable Name | Description | Source |
|---|---|---|
| State | List of all 36 states in Nigeria and the FCT | |
| Geopolitical zones | List of the six (6) geopolitical zones where each state and FCT falls | |
| Longitude, Latitude | A geographic coordinate of each state in Nigeria | |
| No. of Cases (Lab Confirmed) | Cumulative of COVID-19 lab-confirmed cases per state | [3] |
| No. of Cases (On Admission) | Cumulative of COVID-19 cases on admission per state | [3] |
| No. Discharged | Cumulative of COVID-19 discharged cases per state | [3] |
| No. of Deaths | Cumulative of COVID-19 death cases per state | [3] |
| Discharge rate | Proportion of COVID-19 discharged cases and lab-confirmed cases per state expressed in percentage | [3] |
| Fatality rate | Proportion of COVID-19 deaths cases and lab-confirmed cases expressed in percentage | [3] |
| COVID-19 government laboratories | Total government COVID-19 testing laboratories for each state in Nigeria | [3] |
| Healthcare facilities | Total primary, secondary and tertiary healthcare facilities available for each state in Nigeria | [6] |
| 2020 initial budget (bn) presented | Initial budget presented for the year 2020 by the state government before COVID-19 | [5] |
| 2020 revised budget (bn) due to COVID-19 | Revised budget for the year 2020 by the state government due to COVID-19 | [5] |
| Percentage budget reduction | The percentage of budget reduction for each state in Nigeria | [5] |
| Population estimate (0-100) years | Population estimate by gender for each state in Nigeria | [8] |
| Date | Datetime of NCDC tweet | [4] |
| Tweet | Tweets posted or retweet by NCDC | [4] |
| Tweet_type | Whether the tweet is emanating from NCDC directly or retweet | [4] |
| Hashtags | Hashtags included in the tweet | [4] |
| Media_url | Images included in the tweet | [4] |
| Mentions_twitter_handle | People mentioned/tagged in the tweet | [4] |

The first 16 variables in Table 1 are categorized into six (6) geopolitical zones in Nigeria while the features such as discharge rate, fatality rate, and percentage budget reduction were derived by the authors.

Imo), South-South (Akwa Ibom, Bayelsa, Cross River, Rivers, Delta, Edo) and South-West (Ekiti, Lagos, Ogun, Ondo, Osun, Oyo).

**Sheet 4** This sheet shows the general COVID-19 testing laboratories in Nigeria.

**Sheet 5** This sheet contains NCDC daily tweets from December 1, 2019 to September 29, 2020.

The datasets were collected through web scraping of different sources with features referenced and described in Table 1. They contain the daily updates of confirmed, recovered and death cases of COVID-19 in Nigeria [4], cumulative laboratory confirmed cases, patients on admission, number of patients discharged, number of death cases, discharge rate and fatality rate of COVID-19 per state in Nigeria [3], COVID-19 Government laboratories [3], total healthcare facilities per state in Nigeria [6], initial, revised, and percentage budget reduction due to COVID-19 [5] total revenue available per state in 2019 [7], and the population estimate per state from 2016 to 2017 [8].

### 1.1. Nigeria population estimate and healthcare facilities

The Nigeria population estimate per state and the FCT in Fig. 2 was generated from the data that was produced by the WorldPop Research Group at the University of Southampton [8] and it represented bottom-up gridded population estimates (∼100 m grid cells) from 2016 to 2017.

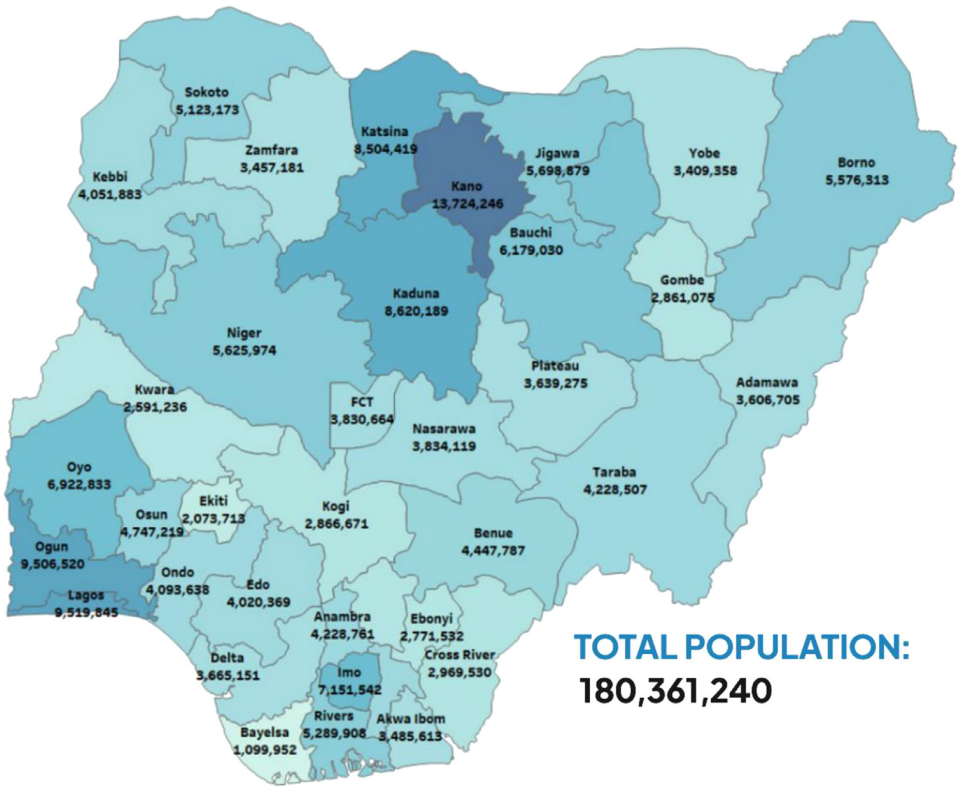

**Fig. 2.** Nigerian states population estimate from 2016 to 2017.

The data included the estimates of the number of people belonging to individual age-sex groups but only total population estimates for all gender was shown in Fig. 2.

The Fig. 3 shows the total primary, secondary, and tertiary healthcare facilities available in Nigeria.

### 1.2. COVID-19 data analytics

#### 1.2.1. COVID-19 cases by geopolitical zone

Fig. 4 shows the distribution of COVID-19 for each geopolitical zone. It can be seen that COVID-19 affected the South-West region more when compared to other zones in Nigeria.

#### 1.2.2. COVID-19 laboratories in Nigeria

To contain the spread of COVID-19 in Nigeria, the government of each state has taken the decision in line with the NCDC directives to equip laboratories and provide testing kits for COVID-19. The government laboratories were spread across the six (6) geopolitical zones and they are as follows: North-East (7, 10.4%), South-East (8, 11.9%), North-Central (15, 22.4%), South-West (15, 22.4%), North-West (12, 17.9%) and South-South (10, 14.9%) as shown in Fig. 5.

Figs. 6 and 7 show different kinds of COVID-19 testing laboratories at which anyone can test for COVID-19 in Nigeria and they include government laboratories, fee paying private laboratories, and corporate laboratories. Corporate laboratories are owned by the corporate bodies such as Shell Petroleum Development Company of Nigeria. As reported by NCDC, a laboratory can

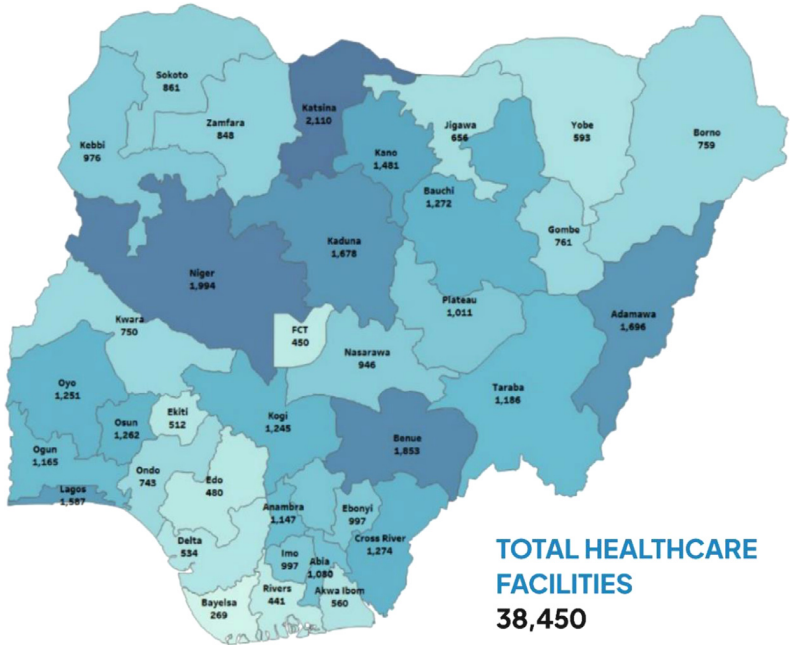

**Fig. 3.** Healthcare facilities available for each state in Nigeria.

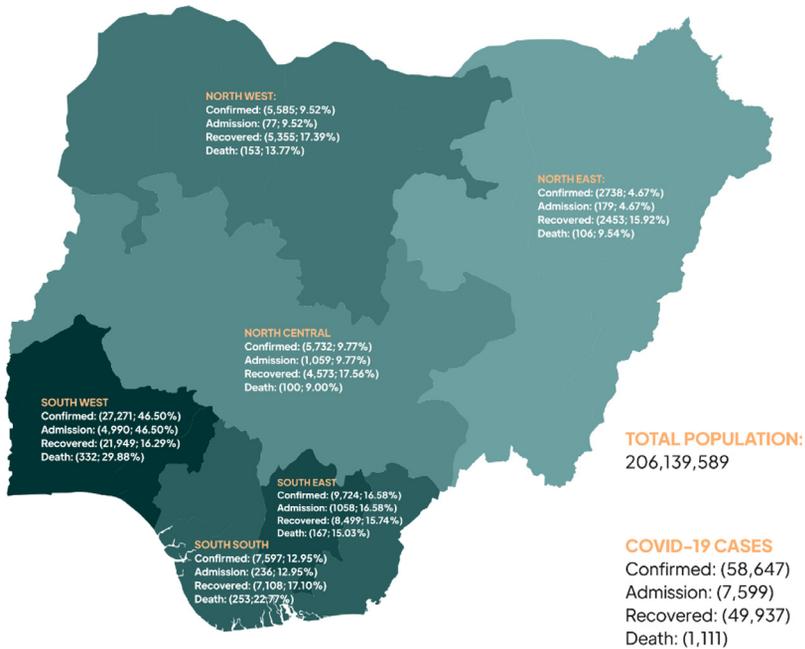

**Fig. 4.** COVID-19 cases as of September 29, 2020 by geopolitical zones and Nigeria 2020 population estimate at mid-year by the UN.

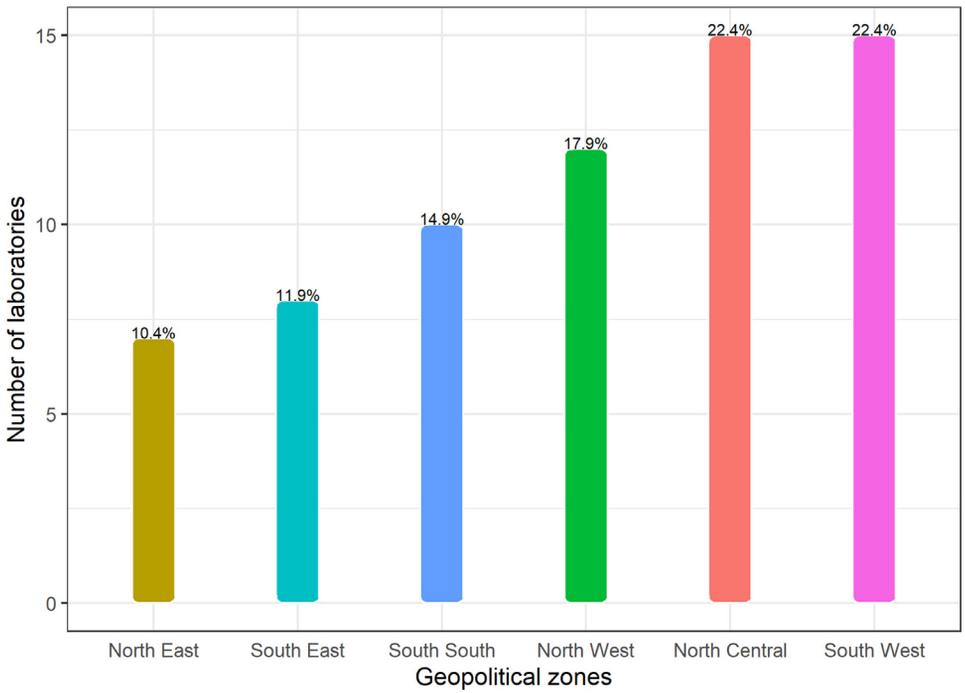

**Fig. 5.** Government Laboratories by Geopolitical Zones as of September 29, 2020.

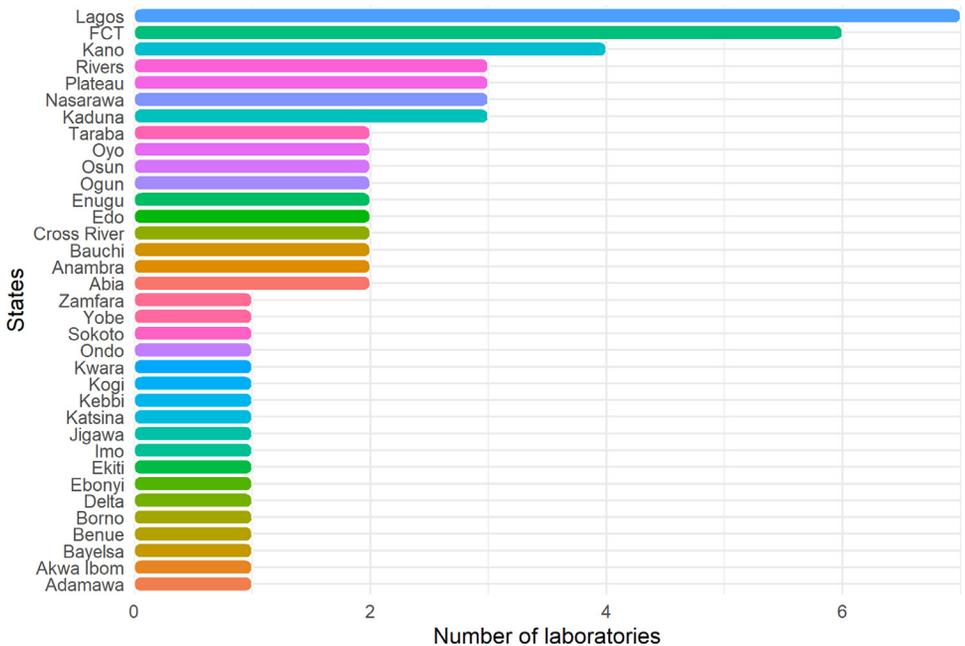

**Fig. 6.** Government laboratories for testing COVID-19 as of September 29, 2020.

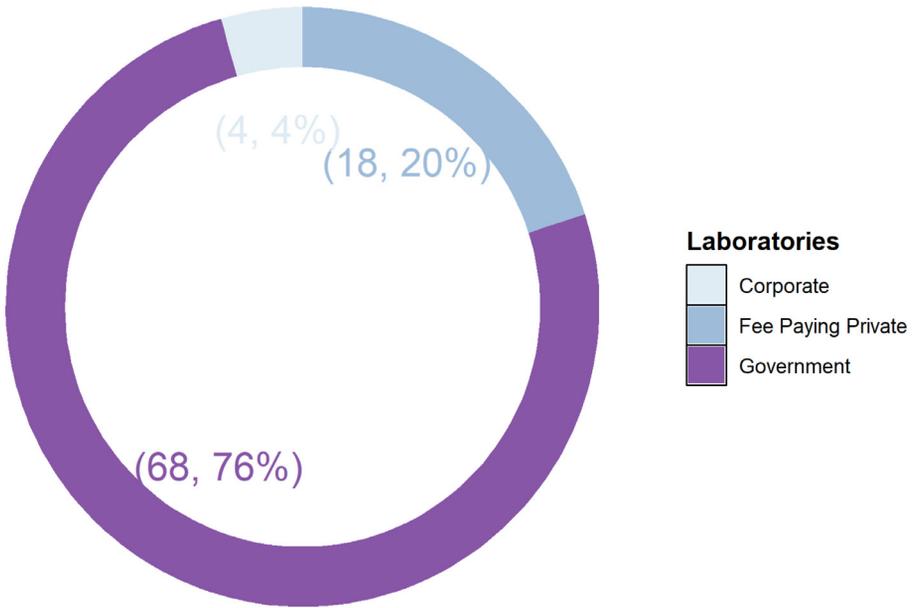

Fig. 7. General COVID-19 testing laboratories as of September 29, 2020.

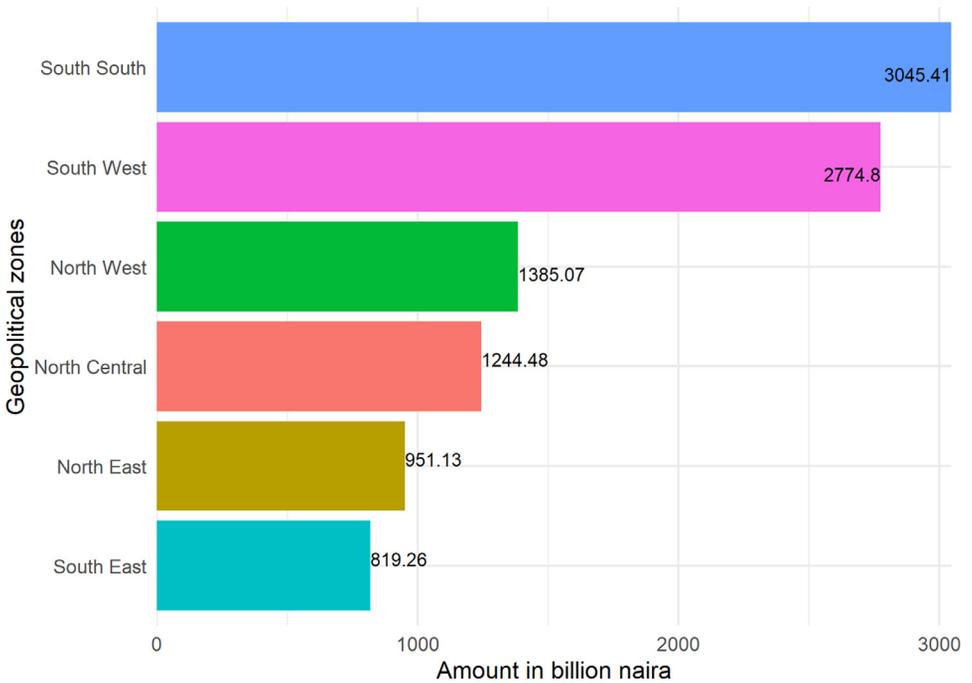

Fig. 8. Initial budget by geopolitical zones in Nigeria.

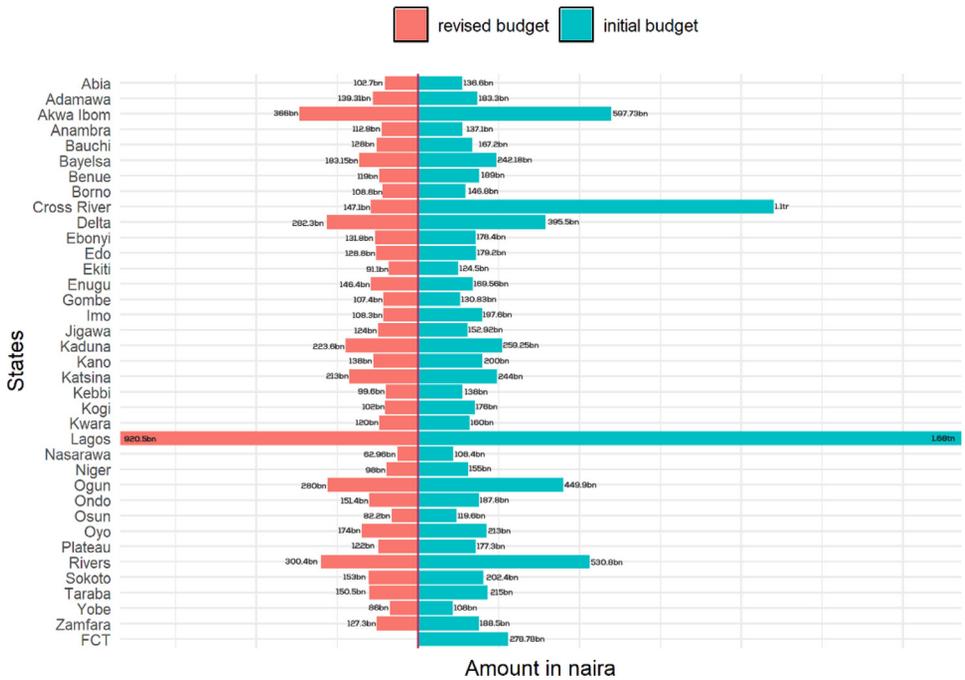

**Fig. 9.** 36 States of Nigeria reduced 2020 budget due to COVID-19 Pandemic.

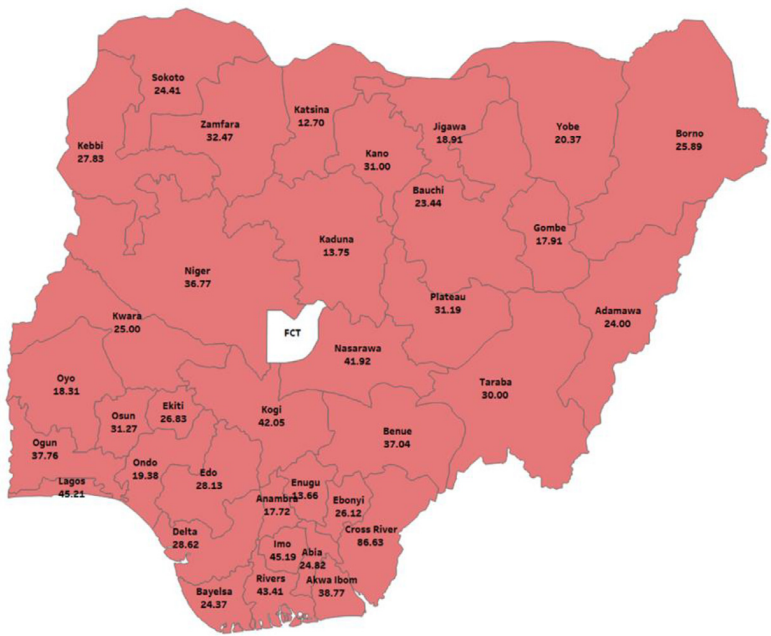

**Fig. 10.** States reduced budget for 2020 due to COVID-19 (Values in percentage).

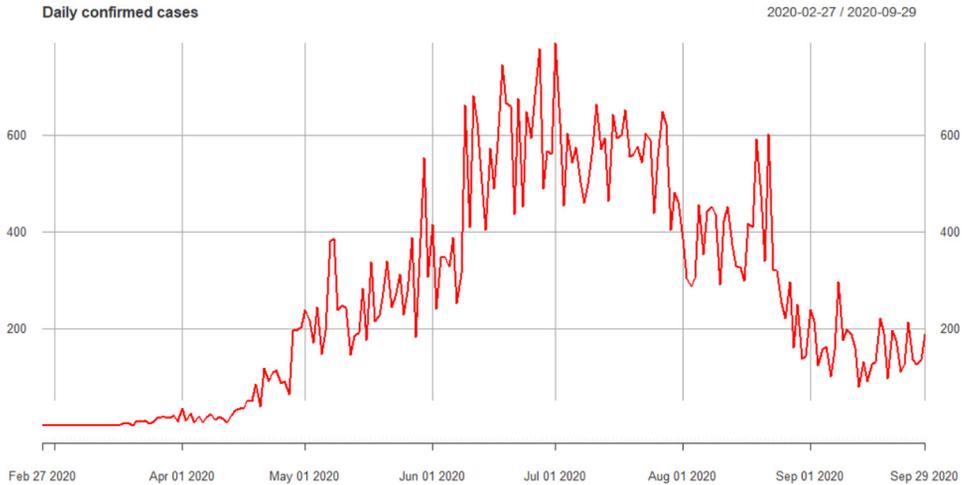

Fig. 11. Daily confirmed cases of COVID-19 in Nigeria.

serve more than one state e.g. African Centre of Excellence for Genomics of Infectious Diseases serves Osun and Ondo States [3].

### 1.3. Budget analysis

Before the incidence of COVID-19 in Nigeria, all the 36 states including the FCT had presented 2020 budgets to their respective House of Assemblies while some had even signed the appropriation bill into law. According to the data that we collated for the initial 2020 states budget, the 36 states and FCT had budgeted a total sum of N10.220 trillion for 2020 compared to the total sum of N9.140 trillion budgeted in 2019. The South-South region with the following states; Cross River, Akwa Ibom, Rivers, Delta, Bayelsa, Edo and in the order of budget ranking has the highest budget of N3.045 trillion combined while the South-West region with the following states; Lagos, Ogun, Oyo, Ekiti, Ondo, Osun and in the order of budget ranking has a total budget of N2.774 trillion as shown in Fig. 8.

All the 36 states excluding the FCT reduced their cumulative budget of N9.941 trillion to N6.131 trillion, making 38.32 % budget reduction due to COVID-19 as shown in Fig. 9.

As shown in Fig. 9, the top five states that have the highest initial budget are Lagos (N1.68 trillion), Cross River (N1.1 trillion), Akwa Ibom (N597.73 billion), Rivers (N530.8 billion) and Ogun (N449.9 billion) while the top 5 states with the highest revised budget due to COVID-19 are Lagos (N920.5 billion), Akwa Ibom (N366 billion), Rivers (N300.4 billion), Delta (N282.3 billion) and Ogun (N280 billion). This suggests that COVID-19 really affected the Cross River state economy forcing them to reduce their budget by 86.63% as shown in Fig. 10 thereby making their initial budget of N1.1 trillion unrealistic.

Also, as shown in Table 2, the South-South region reduced their budget of N3.045 trillion by 53.77% while the South-West region reduced their budget of N2.77 trillion by 38.76%.

## 2. Experimental Design, Materials and Methods

We organized our various datasets using tidy principles; each variable is a column, each observation is a row, and each type of observational unit is a table [9]. Tidy datasets are always easy to manipulate and visualize. Majority of the data wrangling was done in R programming

**Table 2**
Initial and revised 2020 budget by geopolitical zones in Nigeria (36 states + FCT).

| Geopolitical zones | Cumulative Initial budget | Cumulative Revised budget | % reduction |
| --- | --- | --- | --- |
| North-Central | 1,244.48 | 623.96 | 49.86% |
| North-East | 951.13 | 720.01 | 24.30% |
| North-West | 1,385.07 | 1,078.5 | 22.13% |
| South-East | 819.26 | 602 | 26.52% |
| South-South | 3,045.41 | 1,407.75 | 53.77% |
| South-West | 2,774.8 | 1,699.2 | 38.76% |
| Grand total | 10,220.15 | 6,131.42 | 40.01% |

Budget in billion naira.

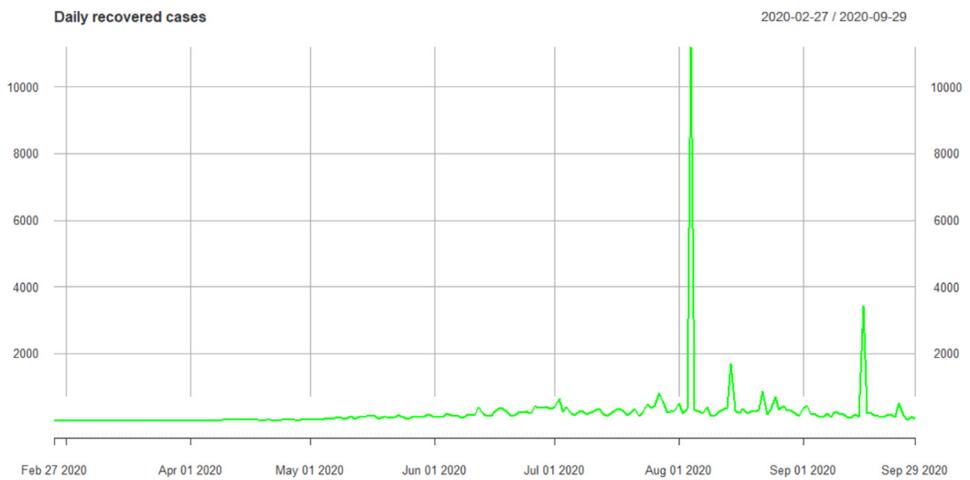

**Fig. 12.** Daily recovered cases of COVID-19 in Nigeria.

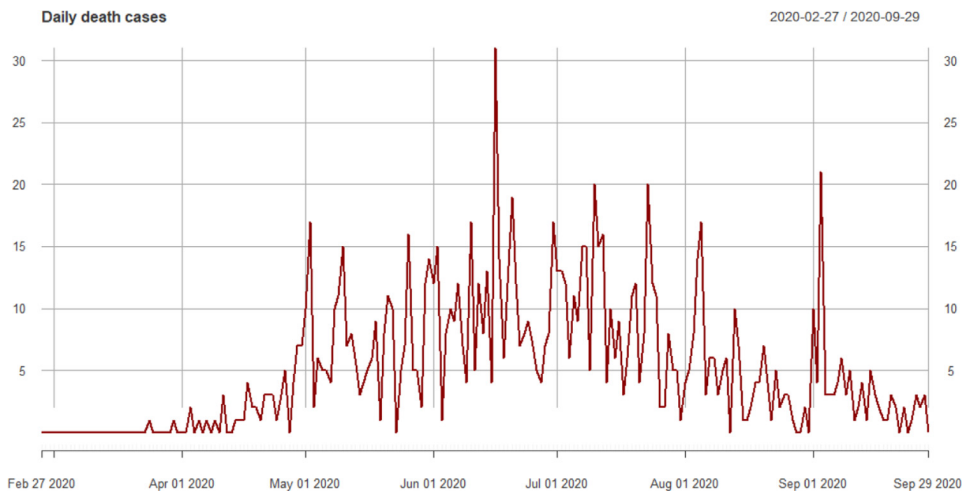

**Fig. 13.** Daily death cases of COVID-19 in Nigeria.

Fig. 14. The NCDC daily tweets and retweets from December 1, 2019 to September 29, 2020.

Fig. 15. Word cloud of NCDC tweets from December 1, 2019 to September 29, 2020.

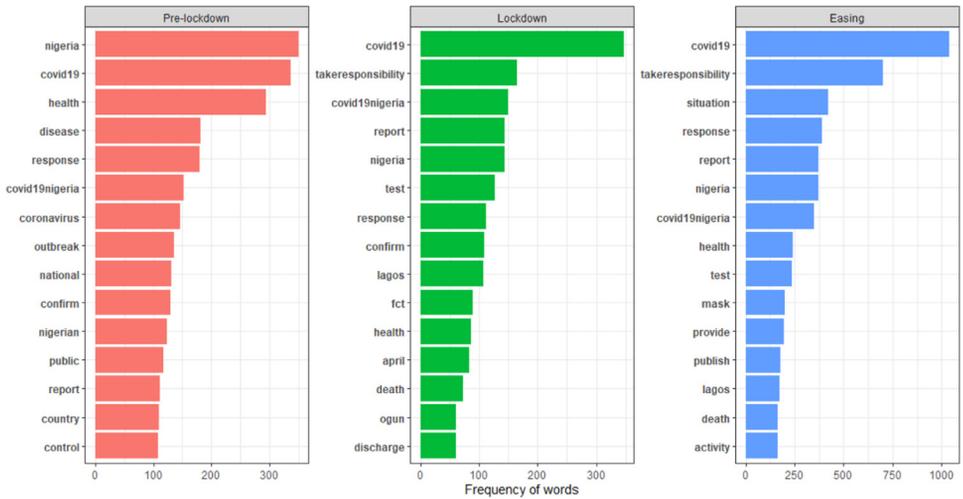

**Fig. 16.** Top 15 words found in NCDC tweets at different COVID-19 phases in Nigeria.

version 4.0.2 while the dataset resides in the Microsoft Excel 2016. Packages used in R included dplyr and ggplot2 for data analysis and visualization [10,11]. We also used the readxl package [12] to import the Excel file to R programming and the janitor package [13] to clean the data. We used xts package [14] to create an extensible time-series object which can be ordered by time index to plot the daily updates of COVID-19 confirmed, recovered and death cases in Nigeria as shown in Figs. 11 to 13. Dataon NCDC timelines on Twitter was scrapped using rtweet [15] and the frequency of daily tweets and retweets from December 1, 2019 to September 29, 2020 is shown in Fig. 14 while most of the common words found in the tweets are shown in Fig. 15.

Fig. 16 shows the frequency of most common words found in NCDC tweets at various COVID-19 phases in Nigeria which were partitioned into pre-lockdown (December 1, 2020 to March 29, 2020), lockdown (March 30, 2020 to May 4, 2020), and lockdown easing (May 5, 2020 to September 29, 2020) to provide a deep understanding and variation in the importance of the mentioned words in Fig. 15. For example, it could be seen that the word 'takeresponsibility' is more frequently used during the lockdown and easing phases.

We carried out most of the data visualization in RStudio and also used Tableau and Corel-DRAW to bring out those visualizations into perfect shape. RStudio project script can be found via https://bit.ly/COVID-19data_project_repo.

## Ethics Statement

The process of data collection does not violate any social media privacy, involve experiment, or human sample.

## Declaration of Competing Interest

This research is supported by Data Science Nigeria (DSN), 174b, Muritala Mohammed way, Yaba, Lagos, Nigeria, https://www.datasciencenigeria.org.

**Acknowledgments**

Our gratitude goes to the Virus Outbreak and Disease Network (VODAN) Africa for the training and workshops in creating the science of FAIR data in Africa. We also sincerely appreciate the Nigerian team of VODAN-Africa, Data Science Nigeria (DSN), and Olabisi Onabanjo University for providing data stewardship, and Oniyitan Oluwaseyi for visualizing Fig. 4 using CorelDRAW. We also thank Wole Ademola Adewole, the Implementation Coordinator of GRID3 Nigeria for his time and support in providing all relevant information about GRID3 data that were used in this data article.

**Supplementary Materials**

Supplementary material associated with this article can be found in the online version at doi:10.1016/j.dib.2020.106424.